\documentclass[runningheads]{llncs}
\usepackage[T1]{fontenc}
\usepackage{mathtools}
\usepackage{amsmath, amssymb}
\usepackage{stmaryrd}
\usepackage{paralist}
\usepackage{cleveref}
\usepackage{algorithm}
\usepackage{listings}
\usepackage{graphicx}
\usepackage{multirow}
\usepackage{booktabs}
\usepackage{paralist}

\let\cref\Cref

\DeclareMathOperator{\pathstart}{start}
\DeclareMathOperator{\pathend}{end}

\DeclareMathOperator{\argmin}{argmin}

\DeclareMathOperator{\prom}{prom}
\DeclareMathOperator{\lp}{LP}
\DeclareMathOperator{\rng}{RNG}
\DeclareMathOperator{\SP}{SP}

\DeclareMathOperator{\mg}{MG}
\DeclareMathOperator{\PG}{PG}

\begin{document}
\title{The Mont Blanc of Twitter: Identifying Hierarchies of Outstanding Peaks
in Social Networks}

\title{The Mont Blanc of Twitter: Identifying Hierarchies of Outstanding Peaks
  in Social Networks}

\author{Maximilian Stubbemann \inst{1}\orcidID{0000-0003-1579-1151}
  \and Gerd Stumme \inst{1}\orcidID{0000-0002-0570-7908}}
\institute{Knowledge and Data Engineering Group, University of Kassel, Kassel,
  Germany
  \email{\{stubbemann stumme\}@cs.uni-kassel.de}}
\titlerunning{The Mont Blanc of Twitter}
\toctitle{The Mont Blanc of Twitter: Identifying Hierarchies of Outstanding Peaks
  in Social Networks}
\authorrunning{Maximilian Stubbemann and Gerd Stumme}
\tocauthor{Maximilian Stubbemann and Gerd Stumme}

\maketitle              
\begin{abstract}
  The investigation of social networks is often hindered by their size
  as such networks often consist of at least thousands of vertices and
  edges. Hence, it is of major interest to derive compact structures
  that represent important connections of the original network. In
  this work, we derive such structures with orometric methods that are
  originally designed to identify outstanding mountain peaks and
  relationships between them. By adapting these methods to social
  networks, it is possible to derive family trees of important
  vertices. Our approach consists of two steps. We first apply a novel
  method for discarding edges that stand for weak connections. This is
  done such that the connectivity of the network is preserved. Then,
  we identify the important ``peaks'' in the network and the ``key
  cols'', i.e., the lower points that connect them. This gives us a
  compact network that displays which peaks are connected through
  which cols. Thus, a natural hierarchy on the peaks arises by the
  question which higher peak comes behind the col, yielding to chains
  of peaks with increasing heights. The resulting ``line parent
  hierarchy'' displays dominance relations between important
  vertices. We show that networks with hundreds or thousands of edges
  can be condensed to a small set of vertices and key connections between them.
  
\end{abstract}

\section{Introduction}
Relationships in social networks are usually modelled as
graphs. Examples of this are follower relations on Twitter or
friendships on Facebook.  However, even for medium-sized graphs with thousands
of nodes to display and comprehend the full structure is often not possible. Another
problem is that the importance of different edges often
varies.
This is especially possible in networks that arise as projections from
other graphs. Examples for this are  networks of co-group memberships of
Youtube users or co-Follower networks on Twitter. Here, there will be a large amount of
``weak'' edges where the set of shared neighbors in the original graph
was small. In such cases, it is crucial to derive compact representations of
structurally important relationships.

Often, the importance of individual vertices can be measured by a given
``height'' function. For example, Twitter users can be evaluated by the amount
of followers and academic authors by their h-index. While it is intuitive to
sample the ``top $k$'' users as a subset, this may not lead to a reasonable
representation of the important nodes. This is for example the case if Twitter
users with high follower counts are surrounded by users with even higher counts.
Hence, they may have a overall large height which is however not outstanding for
the specific community they belong to. In contrast to just assume the
``highest'' vertices as important, we propose a way to identify \emph{locally
  outstanding} nodes in networks, i.e., nodes with a large height with respect to
their surrounding community. Additionally, we derive hierarchical relations
between these outstanding nodes.

Our approach adopts notions from the realm of orometry which are originally
designed to evaluate the outstandingness of mountains. The
\emph{(topographic) prominence} of a mountain quantifies its local
outstandingness by computing the minimal vertical descent that is needed to
reach a higher peak. Paths with minimal descent to a higher peak
deliver two important reference points for each mountain. First, the lowest
point of this path determines the prominence value. This point is called
the \emph{key col}.  Secondly, the first higher peak reached after the
key col is called the \emph{line parent}. Adopting these notions to
networks allows to find locally outstanding nodes and to derive a
compact tree structure which displays how these outstanding nodes are
dominated by each other. 

When deriving such structures, the question arises on how to traverse
the network to find key cols and line parents. Here, it is natural to use the
edges of the graph. However, as mentioned above, some edges in the graph may
represent weak connections and should not contribute to the derived landscape.
Hence, it can be beneficial to remove edges as a preprocessing step. To this
point, we propose a method for parameter-free edge-reducing  based on the
\emph{relative neighborhood graph }(RNG)~\cite{toussaint80}. We will show that
our edge-reduction technique preserves connectivity. This is not guaranteed by
other approaches which discard edges via a weight threshold or only keep the $k$
most important edges. Note, that the key contribution of our approach are mountain
graphs and line parent trees. Discarding unimportant edges is an optional
preprocessing step.

To sum up, our approach derives line-parent trees between locally outstanding
nodes. This significantly simplifies the study of networks because trees can be
satisfactory visualized and navigating through them is possible for larger node
sets. Furthermore, the derived hierarchy is not a subset ot the original edge
relation. Thus, we create a novel view on social networks which is not captured
by existing approaches. We provide our code the sake of
reproducibility.\footnote{\url{https://github.com/mstubbemann/mont-blanc-of-twitter}}

\section{Related Work}
Deriving compact structures that display important
relations in the original network is often done via sampling
vertices or edges~\cite{rafiei05,krishamurthy03,leskovec06,li15,li19}. In contrast, other works focus on the aggregation
of vertices and edges such that the original network can be
reconstructed~\cite{toivonen11,li19a,royer08}.  All these methods have in common that they return a proxy of the original network. Thus, they
are not able to identify hierarchies and connections of outstanding
vertices that are not approximations or explicit subgraphs of the original graph.

The study of hierarchic structures has gained recent interest. Lu et
al.~\cite{lu16} derives acyclic graphs by removing cycles. Other works use
likelihoods to derive suitable hierarchies~\cite{clauset08,maiya09} or provide a
quantification on how ``hierarchical'' a graph is~\cite{gupte11}. In contrast to
our approach, these methods are solely based on the graph structure and are not
able to incorporate the ``height'' of nodes. The usage of the height
function is a unique feature of our line-parent hierarchy, resulting in trees
structure that capture different connections than existing approaches.

The idea of adapting methods from orometry to different areas has been followed
in recent works~\cite{nelson19,karatzoglu20}. On the other hand, there is
a variety of works which study prominence in different abstract
settings~\cite{schmidt19,stubbemann20,pavlik15}. All these works have in common,
that they focus on the computation of prominence. In the present work, we go a
step further and study the underlying structure, i.e., the connections to key
cols and line parents which determine prominence values.

\section{Mountain Graphs and Line Parent Trees}

In this section we present our approach to derive small hierarchies between
peaks from larger networks. We first explain how one can derive mountain graphs
and line parents from networks that provide distance and height information.
Afterwards, we propose an optional preprocessing step that uses the notion of
relative neighborhood graphs (RNGs)~\cite{toussaint80} to remove a significant
amount of edges while preserving the connectivity of the original network. This
will provide us with an end-to-end pipeline for extracting line parent
hierarchies from networks by first  discarding unimportant edges, which is
described in~\cref{sec:rng} and by secondly computing the mountain graph and
line parent tree from the resulting network as described in~\cref{sec:mg}
and~\cref{sec:line_trees}. We rely on the prominence term from Schmidt and
Stumme~\cite{schmidt19}. A complete example of the procedure which we develop in
the following is given by~\cref{fig:subs}. The proofs of all theorems presented
in the following can be found in the supplementary material.\footnote{\url{https://github.com/mstubbemann/mont-blanc-of-twitter}}

\subsection{Landscapes and Mountain Graphs}
\label{sec:mg}

\begin{figure*}[t]
  \centering
  \includegraphics[width=\linewidth]{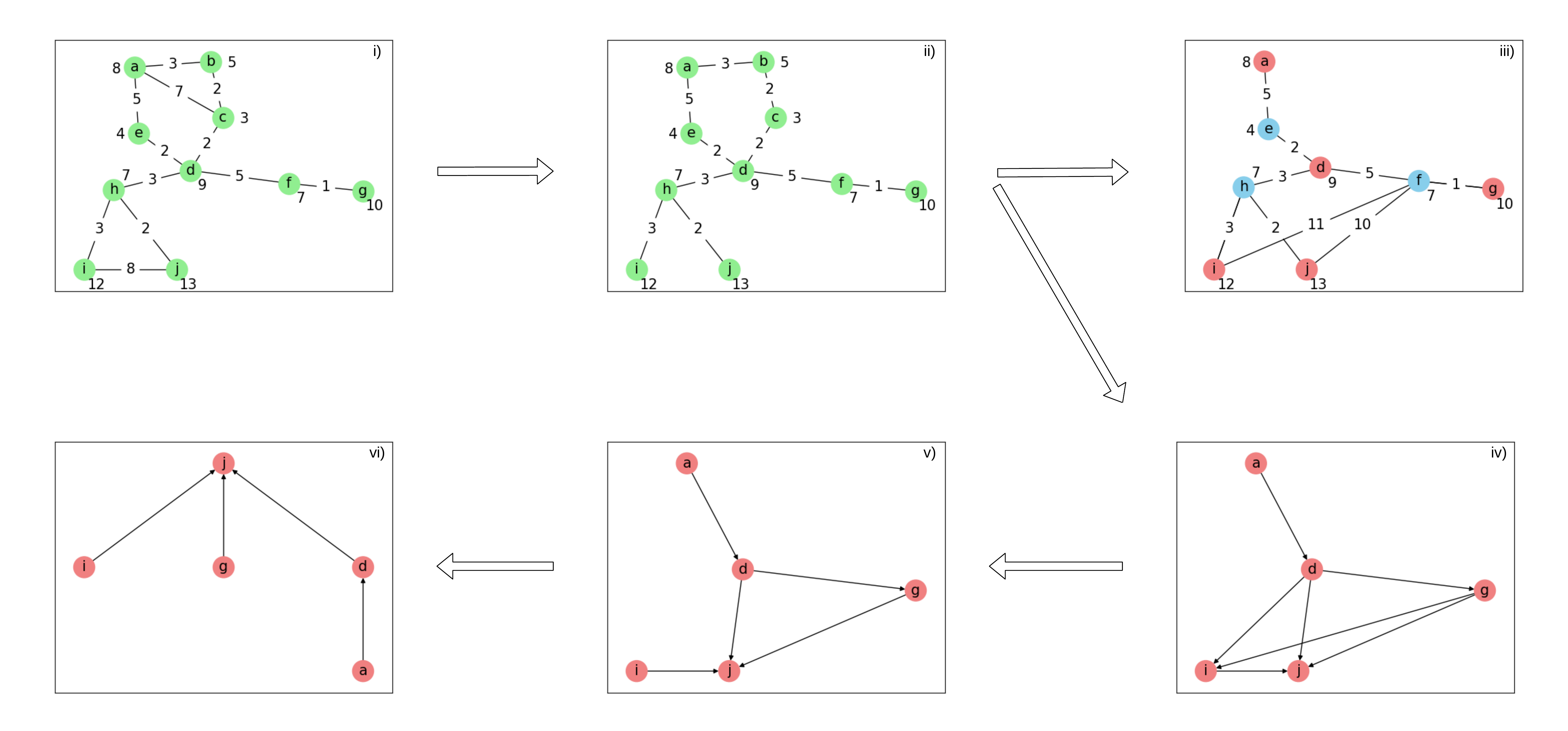}
  \caption{Generating the mountain graph
    and the line parent tree. In i),
    a graph is displayed with the heights next to the nodes and
    with the edge weight put in the middle of each edge. The RNG in ii) is derived by
    discarding the edge between i and j, because h is closer to both i
    and j than they are to each other and by discarding the edge between
    a and c because node b is closer to both of them than they are to
    each other. From this, the Mountain Graph is derived in iii), where we
    display the shortest path distances between the cols and
    peaks. Additionally, we derive from ii) the Peak Graph which is
    displayed in iv). According to \cref{def:LP}, we then discard the edge
    from g to i because j is closer to the key col f than i and we
    discard the edge between d and i because j is closer to the key col  h to arrive at
    v). From this, the line parent tree vi) is derived by discarding the edge
    between d and g because the key col h over which j is reached is
    closer to g than the key col f over which g is reached.}
  \label{fig:subs}
\end{figure*}

In the following, we will work with undirected graphs $G=(V,E)$, where
$E \subseteq \binom{V}{2}$. We call a function
$w\colon E \to \mathbb{R}_{>0}$ a \emph{weighting function} of
$(V,E)$. If we have a triple $G=(V,e,w)$ where $(V,E)$ is an
undirected graph and $w$ a weighting function on $(V,E)$, we call $G$
a \emph{weighted} graph. If we simply speak of graphs, we refer to undirected and
unweighted graphs.

A \emph{walk} $p$ of a graph
$G$ is a finite sequence $p=(v_i)_{i=0}^n$ with $v_i \in V$ for all
$i \in \{0, \dots n\}$ and $\{v_{j-1},v_{j}\} \in E$ for all $j \in \{1,\dots,n\}$. We call a walk
$p$ a \emph{path}, if for all $i \neq j$ it holds that $v_i \neq v_j$ or $\{i, j\}
  = \{1, n\}$. For each
walk $p= (v)_{i=0}^n$, we call $\pathstart(p)\coloneqq v_0$ the \emph{starting
  point}  and $\pathend(p)\coloneqq v_n$ the \emph{end point} of $p$. We follow
the usual convention to not distinguish between walks
$p=(v_i)_{i=0}^n$ and the corresponding set $\{v_i~|~i \in \{0,\dots
  n\}\}$, meaning that we say that $v$ is element of $p$ and write $v
  \in p$. A graph is \emph{connected}, if for all pairs $u,v \in
  V$ there is a walk from $u$ to $v$. Additionally, let $N_G(v)\coloneqq \{u \in V~|~ \{u,v\}
  \in E\}$ be the neighborhood of $v$ in $G$. If clear from
the context, we omit $G$ and write $N(v)$. 

We consider graphs with a \emph{height function} $h\colon
  V \to \mathbb{R}_{\geq 0}$ and a \emph{metric}, i.e., $d \colon V \times V \to \mathbb{R}_{\geq 0}$
with
\begin{inparaitem}
  \item[$\bullet$] $\forall x,y \in V:d(x,y)=0\iff x=y$ (reflexivity), 
  \item[$\bullet$] $\forall x,y \in V:d(x,y)=d(y,x)$ (symmetry)  and
  \item[$\bullet$] $\forall x,y,z \in V: d(x,z) \leq d(x,y) + d(y,z)$ (triangle inequality).
\end{inparaitem}

\begin{definition}[Landscape]
  We call $L=(G,d,h)$ a \emph{landscape} if $G=(V,E)$ is
  a connected\footnote{This is assumed for
    simplicity. The following foundations can be applied to
    unconnected graphs by studying every connected component for itself.} and finite graph, $d$ is a metric on $V$ and $h$ is a height
  function on $V$ such that $h$
  has a unique maximum. We denote the highest point by $\max(L)$.
\end{definition}

To sum up, a landscape is given by a set of points, where we can
traverse the points (via the given graph structure),
where we know, how ``high'' each point is and where we can measure
distances. If $G$ has a weighting function $w$ on it, a metric on the
nodes of $G$ is
provided by the weighted shortest path distance.

As mentioned earlier, our aim is to display hierarchies of peaks and connections
between peaks and cols. For this, we first have to define peaks and cols. In the
following, we will always assume to have given a landscape $L=(G,d,h)$.

\begin{definition}[Peaks, Mountain paths and Cols\footnote{To simplify
      notations, our definition of cols allow only one col per path
      which differs from the definition in geography,}]
  We call a node $v \in V$ a \emph{peak} of $L$
  if $h(v) > h(u)$ for all $u \in N(v)$ and denote by
  $P(L)$ the set of peaks of $L$. A path $p$ of
  $G$ is a \emph{mountain path} if $\pathstart(p),
    \pathend(p) \in P(L)$. We denote by $M(L)$ the set of
  all mountain paths. For each $p \in M(L)$ we call $c(p)\coloneqq\argmin_{v \in p}
    h(p)$ the \emph{col} of $p$. If this argmin is not unique, we choose
  the point in the path which is visited first.
\end{definition}

To compute the prominence of a peak, we have to identify the
cols which connect it with higher peaks.

\begin{definition}[Cols of Peaks]
  For each peak $v \in P(L) \setminus \{\max(L)\}$ we call the set $\uparrow_L(v)
    \coloneqq \{p \in M(L)~|~ \pathstart(p)=v, h(\pathend(p)) > h(v),
    \not \exists u \in p\setminus\{\pathend(p), v\}: u \in P(L) \land h(u) > h(v)\}$
  the \emph{ascending paths} of $v$ and denote by $C_L(v) \coloneqq
    \{c(p)~|~ p \in \uparrow_L(v)\}$ the set of all \emph{cols of $v$}.
\end{definition}

To sum up, the ascending paths $p \in \uparrow_L(v)$ are the paths from $v$ to higher
peaks such that there is no higher peak $w \in p$ and the cols of $v$ are the
lowest points of the ascending paths
We omit the $L$ in the index if clear from the context.
As prominence for mountain peaks is the \emph{minimal} descent
needed to go to higher points, we are just interested in the
\emph{highest} cols.

\begin{figure*}[t]
  \centering
  \includegraphics[width=\linewidth]{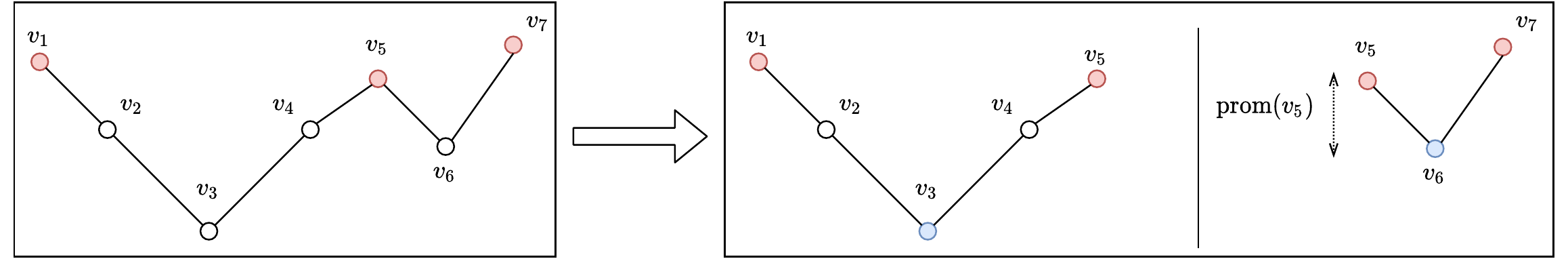}
  \caption{Prominence: Here, the vertical positioning displays the
    height of the different points. To compute the prominence of $v_5$, we first
    identify the paths that lead to higher peaks. Then, we determine
    the cols of these paths and compute the height difference of
    $v_5$ to these cols. The lower difference yield the prominence
    of $v_5$.}
  \label{fig:prom}
\end{figure*}

\begin{definition}[Key Cols and Prominence]
  Let $\prom_L(\max(L))\coloneqq h(\max(L))$. For each peak
  $v \in P(L)\setminus\{\max(L)\}$ we call the elements of
  $K_L(v)\coloneqq \{u \in C_L(v)~|~ h(u)= \max_{\tilde{u} \in
      C_L(v)}h(\tilde{u})\}$ the key cols of $v$. For $u \in K_L(v)$ the
  prominence of $v$ is given via
  \begin{equation*}
    \prom_L(v)\coloneqq h(v) - h(u).
  \end{equation*}
\end{definition}

Thus, the prominence of a peak $v$ displays the vertical distance to the key
cols. For $v \in P(L) \setminus \{\max(L)\}$ the prominence is the minimal
height difference to a col, i.e., $\prom_L(v)=\min_{u \in C(v)}h(v) - h(u).$
Again, we write $\prom(v)$ and $K(v)$ if the choice of the landscape is clear.
An illustration of the definition of prominence is given by~\cref{fig:prom}.

We are interested in the structure which determines the prominence of peaks, i.e., in the
higher peaks to reach from a specific peak and in the cols
which connects the peaks of the mountain landscape. Hence, we do not only study
the key cols of peaks but also the higher peaks that can be reached
from their cols.

\begin{definition}[Dominators]
Let $v$ be a peak of the landscape $L$. We then call the set $D_L(v)\coloneqq
\{\pathend(p)~|~p \in \uparrow_L(v)\land c(p) \in K(v)\}$ the \emph{dominators}
of $v$.
\end{definition}

\begin{definition}[Mountain Graph] For a given
  landscape $L=((V,E),d,h)$, let
  $K(L) \coloneqq \cup_{v \in P(L)}K_L(v)$ be the set of \emph{key
    cols of L} and let $V_{\mg}(L)\coloneqq P(L) \cup
    K(L)$ be  the \emph{critical points} of $L$. Let for $v \in P(L)$ be
  $\uparrow_L^K\!(v)\coloneqq \{p \in \uparrow_L\!(v)~|~c(p) \in K_L(v)\}$
  the ascending paths of $v$ with key cols as cols. Let then
  \begin{equation*}
    E_{\mg}(L)\coloneqq \bigcup_{v \in P(L)} \left(\bigcup_{p \in \uparrow_L^K\!(v)}
    \{\{v, c(p)\}, \{c(p), \pathend(p)\}\}\right).
  \end{equation*}
  The graph $\mg=(V_{\mg}(L),E_{\mg}(L))$ is called the
  \emph{mountain graph} of $L$ and the landscape
  $L_{\mg}\coloneqq(\mg_L,d|_{\mg},h|_{\mg})$ the \emph{mountain
    landscape of $L$}.
\end{definition}

To sum up, if a peak $v_1$ is connected via a key col $u$ to a higher
peak $v_2$, we add edges between $v_1$ and $u$ and between $u$ and
$v_2$ to the mountain graph. Thus, the mountain graph displays which
peaks are connected through which key cols. If clear from the context,
we omit $L$ and simply write $\mg=(V_{\mg},E_{\mg})$.  The mountain
graph contains all relevant information for the computation of
prominence values as the following theorem shows.

\begin{theorem}
  \label{theorem:mountain-graph}
  The
  following statements hold:
  \begin{enumerate}
    \item $\mg$ is connected.
    \item $P(L)=P(L_{\mg})$
    \item Consider for each peak $v \in P(L)\setminus \{\max(L)\}$ of $L$ the set
          $N'_{\mg}(v)\coloneqq\{u \in N_{\mg}(v)~|~\exists v' \in
            N_{\mg}(u): h(v') > h(v)\}$. Then:
          \begin{equation*}
            u \in N'_{\mg}(v) \Rightarrow \exists u' \in  C_L(v): h(u') \geq h(u).
          \end{equation*}
    \item It holds for $v \in P(L)\setminus \{\max(L)\}$ that:
          \begin{equation*}
            \prom_L(v) = \min_{u \in N'_{\mg}(v)}(h(v) -h(u)).
          \end{equation*}
  \end{enumerate}
\end{theorem}
\cref{theorem:mountain-graph} shows that to study relations between
cols and peaks that determine prominence values, it is sufficient to
check the cols to which a peak is connected in the mountain
graph. Note, that the key cols and the paths between
peaks passing through them have to be determined to derive the
mountain graph. Hence, \cref{theorem:mountain-graph} does not allow
for a faster computation of prominence values. Instead, it provides a
representation that can be used to observe important
connections between peaks and cols.

\subsection{Line Parent Trees}
\label{sec:line_trees}

As the prominence of mountain peaks is computed by descending to key cols and
then ascending to higher peaks, a hierarchy between peaks arises by the question
to which higher peak one can traverse from a key col of a given peak.

\begin{definition}[Peak Graph]
  Let 
  \begin{equation*}
    E_P(L)\coloneqq \bigcup_{v \in P(L)} \{\{\pathstart(p),
    \pathend(p)\}~|~ p \in \uparrow_L^K\!(v)\}.
  \end{equation*}
  We call
  $\PG(L)\coloneqq (P(L), E_P(L))$ the \emph{peak graph} of $L$ and we
  call
  \begin{equation*}
    T_{\PG(L)}\coloneqq \{(\pathstart(p),
    c(p),\pathend(p))~|~ p \in \uparrow_L^K\!(v)\}
  \end{equation*}
  the \emph{defining
    triples} of $\PG(L)$.
\end{definition}
Peaks may be connected to different key cols and different higher
peaks. To define a meaningful hierarchy on the peaks, we use the
metric $d$ to determine a unique line parent for all  peaks.

\begin{definition}[Line Parents]\label{def:LP}
  Let
  $T'_{\PG}(L)\coloneqq \{(v,u,\tilde{v}) \in T_{\PG} \mid \not\exists 
    v': (v,u,v') \in T_{\PG} \land (d(u,v') < d(u,\tilde{v}) \lor
    (d(u,v') = d(u,\tilde{v}) \land h(v') > h(\tilde{v})))\}$. Let
  $E_{\lp}(L) \coloneqq \{\{v, \tilde{v}\}~|~ \exists u:
    (v,u,\tilde{v}) \in T'_{\PG}(L) \land \not \exists
    u', v': (v,u',v')
    \in T'_{\PG}(L) \land d(v,u') < d(v,u)\}$. We call
  $\lp(L)\coloneqq(P(L),E_{\lp}(L))$ the \emph{line parent
    graph} of $L$. If
  $\{u,v\} \in E_{\lp}(L)$ with $h(v)>h(u)$, $v$ is a \emph{line
    parent} of $u$.
\end{definition}

Again, we omit $L$ when possible without confusion.
In~\cref{def:LP}, we first remove edges to higher peaks that are
further away from the corresponding key col. If there are multiple higher
peaks with the exact same distance to the key col, we keep the highest peak. Then we remove
edges where the key col is further away. If for all
$v \in P(L) \setminus \max(L)$ the line parent is unique, $\lp(L)$ is a
tree.

\begin{theorem}\label{thm:tree}
  If for each peak $v \in P(L) \setminus \{\max(L)\}$ the line
  parent is unique, then $\lp(L)$ is a tree.
\end{theorem}

The uniqueness of the line parent is only violated in two cases. First, if there
are multiple peaks being reached after the same  key col with the exactly same
distance to the key and the same height. In such a case, we can enforce the
uniqueness by sampling one of the peaks. Secondly, if there are  key cols $c_1,
\dots, c_n$ with corresponding higher peaks $p_1\dots p_n$ with the exact same
distance to the point. In such a case, we choose $p_i$ such that $d(c_i,p_i)$ is
minimal. If these minimum is reached multiple times, we enforce uniqueness by
sampling one of the higher peaks with minimal distance to the corresponding key
col.

To sum up, we enforce the uniqueness of the line parent. The simple edge structure of trees enables a
satisfactory visualization even for medium sized node sets.
The line parent tree can also be used to study dominance relationships with a
non peak as a starting point. In this case, we suggest to navigate
through the line parent tree starting with the closest peak with
respect to the given metric.

\subsection{Discarding Edges via Relative Neighborhood Graphs}
\label{sec:rng}

Let  $G=(V,E,w)$ be a weighted graph
and let $h\colon V \to \mathbb{R}_{\geq 0}$ be a height function on $G$. In the following,
we extend this structure to a landscape by using the shortest path
metric on $G$. In practical
applications, the amount of peaks will often be very low. One reason
for this is the huge amount of connections one may have in social
networks. Let us for example assume to have a weighted co-follower
graph (for example weighted with Jaccard-distance) where the height
function is given by the amount of followers. Here, all pairs of users with just one
common follower would be connected and thus nearly all users would
have a ``higher'' neighbor and thus will not be peaks. Hence, it is of
major interest to only keep edges which stand for a strong connection,
i.e., edges between users with a large amount of
common followers.

A straight-forward way to remove edges would be by  choosing a $k \in
\mathbb{N}$ and keep for all vertices only the $k$ edges with the smallest
weights or to choose a $t \in (0,1)$ and remove all edges with weights higher
than $t$. However, besides the disadvantage that in both cases a parameter has
to be chosen, this procedure can lead to disconnected graphs. Restricting to the
biggest connected component of the resulting graph would then lead to the
discarding of whole regions of the graph. To this end, we develop in the
following a parameter free, deterministic edge sampling approach which
always preserves connectivity. This approach is based on the relative
neighborhood graph (RNG)~\cite{toussaint80}. The RNG derives a graph structure
from a metric space by connecting points nearby. More specifically, two points
are connected if there is no third point which is closer to both of them.

\begin{definition}[Relative Neighborhood Graph]
  The \emph{relative neighborhood
    graph} of a metric space $(M,d)$ is
  given by the undirected graph
  $\rng(d)\coloneqq(M,E_{\rng(d)})$ with $E_{\rng(d)}\subseteq
    \binom{M}{2}$ such that $\{m_1, m_2\} \in E_{\rng(d)}$ if and only if
  there does not exist  $m_3 \in M$ with $\max(\{d(m_1, m_3),d(m_2, m_3)\}) < d(m1,m2)$.
\end{definition}

Our goal is to thin out graphs by computing the RNGs. Hence, it is of
fundamental interest that RNGs are connected. For points in $\mathbb{R}^2$, it
has been shown that the RNG is a supergraph of the
minimum-spanning-tree~\cite{toussaint80} which implies
connectivity~\cite{jaromczyk92}. Because RNGs are commonly only studied in
$\mathbb{R}^d$ with $L^p$ metrics we could not find a proof for the connectivity
in arbitrary finite metric spaces. Hence, we prove it in the supplementary
material.

\begin{theorem}[Connectivity of relative neighborhood graph]\label{theorem:con}
  Let $(M,d)$ be a finite metric space. Then $\rng(d)$ is connected.
\end{theorem}

What still needs to be shown is that deriving RNGs from the shortest-path metric
is indeed an edge-reduction technique, i.e., that edges are just removed and
that is not possible that new edges are added.

\begin{theorem}[RNG as Edge-Reduction]\label{theorem:edges}
  Let $G=(V,E,w)$ be a connected, undirected and weighted graph and $d_{\SP}: V
    \times V \to \mathbb{R}_{\geq 0}$ be the shortest path metric on
  $G$. Then it holds that $E_{\rng(d_{\SP})} \subseteq E$.
\end{theorem}

In the following, we use the term \emph{relative neighborhood graph of $G$}, denoted
by $\rng(G)$, which will always refer to the RNG with respect to the
shortest-path-metric. A sketch of an edge-reduction on a graph is
given as part of \cref{fig:subs}.

For a weighted graph $G=(V,E,w)$ with a height function $h$, our standard procedure
is to
\begin{inparaenum}
  \item compute the weighted shortest path metric $d_{\SP}$,
  \item compute $\rng(G)$,
  \item derive from this the following landscape.
\end{inparaenum}

\begin{definition}[Essential Landscape]
  Let $h: V \to R_{\geq 0}$ be a height
  function on a graph $G=(V,E,w)$. Let $d_{\SP}$ be the weighted shortest path metric on $G$. We call
  \begin{equation*}
    L(G,h)\coloneqq(\rng(G), d_{\SP},
    h)
  \end{equation*}
  the \emph{(essential) landscape} of $G$ and $\mg(G,
    h)\coloneqq\mg(L(G,h))$ the \emph{essential mountain graph} of
  $G$. We call
  $\lp(G, h)\coloneqq\lp(L(G,h))$ the \emph{(essential) line parent tree}
  of $G$. If clear from the context, we simply
  write $\mg(G)$ and $\lp(G)$.
\end{definition}

\paragraph{Complexity.}
The naive approach to compute the RNG for a finite metric space $(M,d)$ would be
to check for all pairs  $m_1 \neq m_2 \in M$ whether there exists $m_3$ which is
closer to both of them. This results in an algorithm with runtime
$\mathcal{O}(|M|^3)$~\cite{toussaint80}. For $\mathbb{R}^d$  with a $l_p$
metric, there are algorithms with better
runtime~\cite{toussaint80,agarwal92,jaromczyk92a}. However, these results can
not be applied to shortest path metrics. To compute $\rng(G)$ for a graph
$G=(V,E)$, we can use~\cref{theorem:edges} to speed up the computation as we
only have to check the elements of $E$ and not all node pairs. Hence, computing
the RNG has complexity $\mathcal{O}(|E|||V|)$.

\section{Line Parent Trees of Real-World Networks}

\begin{table}[t]
  \caption{Network statistics. In the first table, we display from left to right: 1.) the number of
    vertices of the network, 2.) its density 3.) the density of the RNG 4.) the
    number of vertices of the mountain graph, 5.) the density of the mountain graph,
    6.) the number of vertices in the line parent tree, 7.) its maximum width and
    8.) its depth. In the second table, we show the node sizes and degrees of the sampled graphs
    serving a) as a comparison for discarding edges via the RNG procedure and b) for
    serving as a comparison for the mountain graph which is computed from the RNG.
    For the latter, we apply the sampling baselines on the RNG, not on the original
    network itself.}
  \centering
  \begin{tabular}[t]{l|lllll|lll}

                 & $|V|$ & $D_G$  & $D_{\rng(G)}$ & $|V_{\mg}|$ & $D_{\mg}$ & $|V_{\lp}|$ & $W_{\lp}$ & $DP_{\lp}$ \\
    \midrule
    Twitter>10K  & 6635  & .9958  & .0005         & 1171        & .1089     & 652
                 & 88    & 20                                                                                      \\
    Twitter>100K & 430   & 1.0000 & .0064         & 146         & .1084     & 84
                 & 14    & 13                                                                                      \\
    ECML/PKDD    & 742   & .0123  & .0052         & 190         & .1957     & 98
                 & 21    & 10                                                                                      \\
    KDD          & 1674  & .0100  & .0036         & 219         & .2236     & 115         & 30        & 8          \\
    PAKDD        & 889   & .0124  & .0054         & 132         & .2155     & 67          & 27        & 5
  \end{tabular}\\
  \vspace{.5cm}
  \begin{tabular}[t]{l|llll|llll}
                 & \multicolumn{4}{l}{$\rng$ Baselines} & \multicolumn{4}{|l}{$\mg$ Baselines}                                                                                                                   \\
                 & \multicolumn{2}{l}{ES}               & \multicolumn{2}{l}{CNARW\cite{li19}} & \multicolumn{2}{|l}{RPN\cite{leskovec06}} & \multicolumn{2}{l}{RCMH\cite{li15}}                                 \\
                 & $|V|$                                & $D_G$                                & $|V|$                                     & $D_G$                               & $|V|$ & $D_G$ & $|V|$ & $D_G$ \\
    \midrule
    Twitter>10K  & 5192.8                               & .0008                                & 5174                                      & .0007                               & 271.1 & .0079 & 1171  & .0021 \\
    Twitter>100K & 374.8                                & .0084                                & 325.5                                     & .0112                               & 35.9  & .0652 & 146   & .0168
    \\
    ECML/PKDD    & 650.7                                & .0067                                & 560                                       & .0092                               & 77.1  & .0304 & 190   & .0153 \\
    KDD          & 1575.5                               & .0041                                & 1407.5                                    & .0041                               & 67.2  & .0390 & 219   & .0138 \\
    PAKDD
                 & 814.7                                & .0064                                & 704.1                                     & .0087                               & 34.3  & .0812 & 132   & .0222
  \end{tabular}
  \label{tab:stats}
\end{table}

We experiment with networks built from a Twitter
follower network~\cite{kwak10,boldi11,boldi04} which we found at
SNAP~\cite{leskovec14} and with networks that display co-author
relations. These networks are derived from the \emph{Semantic Scholar
  Open Research Corpus}~\cite{ammar18}. From the Twitter dataset, we
derive two weighted co-follower networks. In these networks two users
have an edge if they have a common follower. The edges are weighted
via Jaccard distance. We derive a version containing users with at
least $10,000$ followers (\textbf{Twitter>10K}) and a network
containing users with at least $100,000$ followers
(\textbf{Twitter>100K}). Here, the height of a user is given via the
amount of followers.

The co-author networks are derived by considering communities of authors that
regularly publish at a specific conference. We derive datasets for the
\emph{European Conference on Machine Learning and Principles and Practice of
Knowledge Discovery in Databases} (\textbf{ECML/PKDD}), the \emph{SIGKDD
Conference on Knowledge Discovery and Data Mining} (\textbf{KDD}) and the
\emph{Pacific-Asia Conference on Knowledge Discovery and Data Mining}
(\textbf{PAKDD}). The heights of the authors  are given via h-indices.

For all graphs, we use well-established measures to get further insights into
our notions. To be more detailed, we display node sizes and densities of the
networks themselves, the RNGs and the mountain graphs derived from the RNGs.
Additionally, we display node sizes, maximum widths and depths of the
line-parent trees derived from the RNGs. The results can be found
in~\cref{tab:stats}. Plots of all labeled trees and details on dataset creation
are part of the supplementary material.\footnote{\url{https://github.com/mstubbemann/mont-blanc-of-twitter}}

\subsection{Comparison with Sampling Approaches}
\label{sec:comp}

To further understand the steps of our approach, we compare them with commonly
used sampling approaches~\cite{li19,leskovec06,li15}. To be more specific, we
sample edges from the original network to get graphs which have an equal amount
of edges as the RNG. Then we take the biggest connected component of these
graphs.  We call these methods the \textbf{RNG Baselines}. We use two sampling
approaches: First, we sample edges with the probability of an edge e to be
chosen being proportional to $1 - w(e)$, where $w(e)$ is the weight of the
edge\footnote{We use $1 - w(e)$ instead of $w(e)$ because we assume edge weights
to be distances, not similarities.}. We call the resulting baseline the
\emph{Edge Sampling} (ES) approach. As a second comparison, we use a weighted
version of \emph{CNARW}~\cite{li19}, a modern random walk approach.

Additionally, we use sampling approaches to sample from the RNG in such a way,
that we have an equal amount of nodes as in the mountain graph and  take the
biggest component of the resulting network. We call these methods the \textbf{MG
Baseline}. First, we sample nodes by their PageRank value
via\emph{RPN}~\cite{leskovec06}. Again, we also use a modern random walk based
approach, namely \emph{RCMH}~\cite{li15}. The CNARW method used above relies on
common neighbors. Since triangles in the RNG are very uncommon (for these, 2 of
the 3 corresponding edges in the original graph need to have the same distance
weight), we use RCMH instead.

Note, that we use the comparison with other methods to contextualize our novel
structures. As our structures have a different purpose, namely displaying
important connections that are derived from the original network, they are not
directly comparable to regular sampling approaches. These approaches derive
small graphs that behave similar  to the original graph with respect to specific
measures. This makes it unreasonable to interpret the comparison to our
baselines as a competition where higher/lower node sizes or densities are, in
some way, better. As our comparison methods include random sources, we repeat
them 10 times and report means. Statistics, including sizes of the derived RNGs,
mountain graphs and line parent trees, can be found in~\cref{tab:stats}.
Additionally, we include node sizes and densities for all comparison approaches.

We observe that computing the RNG reduces the density by a large
margin. It stands out, that this effect is stronger for the dense
Twitter networks. When sampling an equal amount of edges, there are
nodes which do not belong to the biggest connected component
anymore. This results in a higher density of the (biggest component)
of the networks created by sampling compared to the RNG.

The resulting line parent trees are much smaller than the original
network, reducing the node set by a factor of about $5$ to $10$
times. Another remarkable point is that the mountain graph is always
denser than the RNG from which it is computed and than the graphs
which are sampled via the comparison methods. An explanation is that
edges from a peak $v$ to a col $u$ in the mountain graph correspond to
paths (not edges!) in the RNG. As the amount of paths in a graph is
commonly remarkably higher than the amount of edges, this could be one
reason for the higher density of the mountain graph.

\subsection{Distances to Line-Parent Trees}

\begin{table}[t]
\caption{Mean, median and maximum of the minimal shortest path distance (MSPD)
from all non-peaks $v$ to the set  $P$ of all peaks (left) to the set $H$ which
contains the $|P|$ highest nodes of the network (right).}
\centering  \centering\begin{tabular}{l|lll|lll}
  & $d(P)_{\text{Mean}}$ & $d(P)_{\text{Median}}$ & $d(P)_{\text{Max}}$ & $d(H)_{\text{Mean}}$ & $d(H)_{\text{Median}}$ & $d(H)_{\text{Max}}$ \\
\midrule
Twitter>10K  & 0.87                 & 0.88                   & 1.00                & 0.90                 & 0.91                   & 1.00                \\
Twitter>100k & 0.86                 & 0.87                   & 0.97                & 0.92                 & 0.95                   & 0.99                \\
ECML         & 1.28                 & 1.00                   & 2.98                & 1.74                 & 1.91                   & 4.91                \\
KDD          & 1.30                 & 1.00                   & 2.95                & 1.83                 & 1.95                   & 3.90                \\
PAKDD        & 1.46                 & 1.00                   & 3.78                & 1.94                 & 1.96                   & 4.90
\end{tabular}
  \label{tab:sp}
\end{table}

To investigate to which extent line parent trees are representative for the
structure of the whole network, we compute how ``dense'' the line parent trees
lay in the networks, i. e. the shortest path lengths from all non-peaks to the
peaks. To evaluate whether choosing locally outstanding nodes lead to a better
representation than choosing nodes solely based on their height, we compare our
approach with a ``naive'' approach of assuming the $n$ highest points to be
relevant, where $n$ is the amount of peaks. To be more detailed, we compute for each non-peak $v$ the minimal
shortest path distances (MSPD) to all peaks in the original graph $G$. We do the
same using the $n$ highest points instead of the set of peaks. We report means,
medians and maximum values over the MSPDs of all non-peaks, the results can be found
in~\cref{tab:sp}.

Our results show that locally outstanding nodes better reflect the overall
network then just choosing the highest nodes, with median and mean values of the
MSPDs being fundamentally lower. Furthermore, median MSPDs to peaks are
always not higher then $1$. In contrast, MSPDs to the ``highest'' nodes have median
values of nearly $2$ for the sparse co-author networks. This indicates that
selecting locally outstanding points indeed lead to a more reasonable
representation instead of selecting nodes solely by their hight, ignoring
spatial information. Note, that we compute distances in a weighted
graph. Hence, shortest path distances (and thus medians and maxima over them) do
not have to be integers.

\section{Experiments on Random Data}
\label{sec:studies}
\begin{figure}[t]
  \centering
  \includegraphics[width=\linewidth]{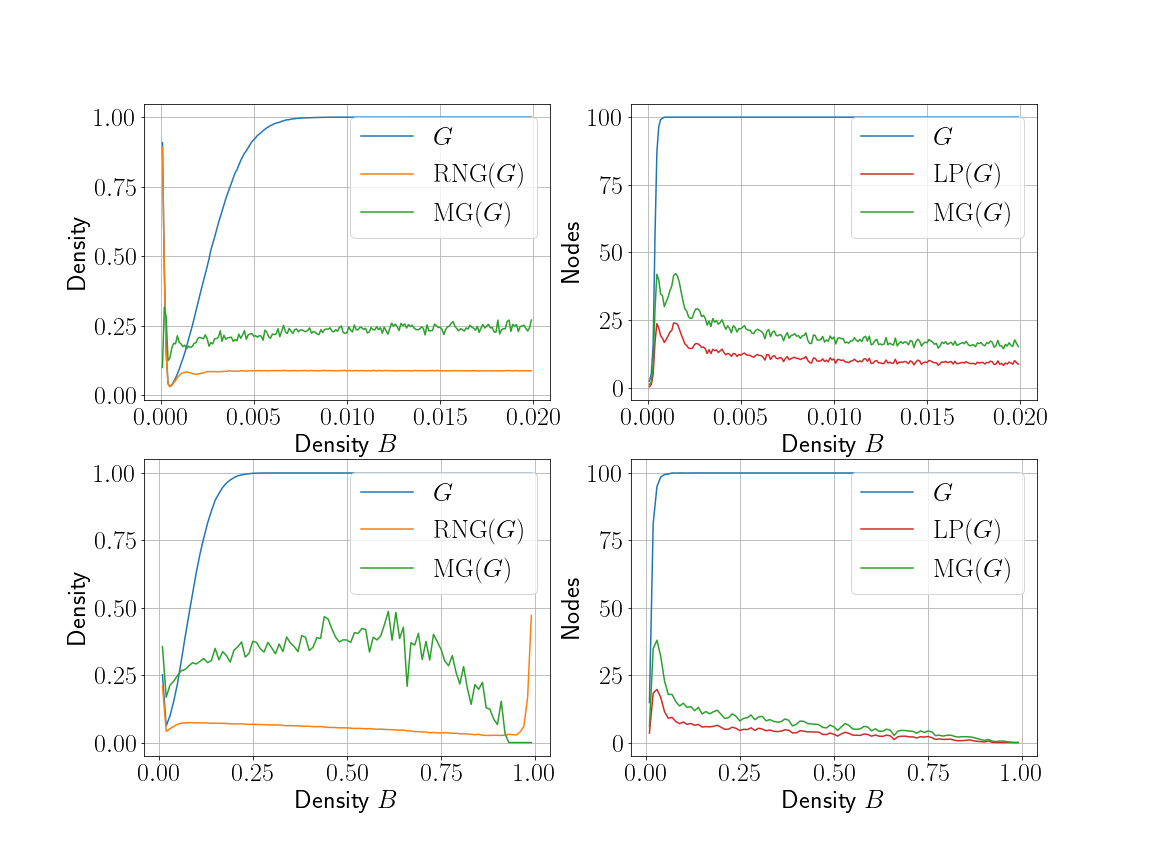}
  \caption{Experiments on random data. In both rows, the set $M_1$ on which is projected
    has size $100$. The other set has size $100,000$ on the first row
    and size $100$ in the second row. The x-axes display the
    densities of the original bipartite graph $B$. The left plots display the
    densities for the resulting network $G$, which is the biggest
    connected component of the weighted projection, the line parent
    tree  $\rng(G)$ and the
    mountain graph $\mg(G)$. The right pictures plot the node size
    of $G, \mg(G)$ and $\lp(G)$.}
  \label{fig:random}
\end{figure}

To investigate sizes and densities of the RNG, the mountain graph and the line
parent tree, we additionally experiment with randomly generated data. Here,  we
start with a randomly generated bipartite graph with vertex sets $M_1,M_2$ with
$|M_1|=100$ and $|M_2| \in \{100000,100\}$. We then project on the vertex set
$M_1$ and set for two vertices with an edge in the resulting graph the edge weight
to the Jaccard distance. The graph $G$ is then given via the biggest connected
component of this graph. As height function, we map each vertex to the amount of
neighbors in the original bipartite network. This procedure is motivated by the
background of often investigated real-world networks. Co-author networks are for
example projection from the bipartite author-publication graph. Here, the
corresponding height function then would be the amount of papers of an author,
where each author is connected to multiple publications but only a small amount
of the overall publications. This leads to a small density of the bipartite
network.

We generate networks for different densities $d$. Namely, we
iterate $d$ through $\{0.0001, 0.0002, \dots, 0.01999\}$ for
$|M_2|=100,000$ and through $\{0.01, 0.02, \dots 0.99\}$ for
$|M_2|=100$.  The experiments with different sizes of $|M_2|$ allow
us to investigate if our methods behave fundamentally different for
networks of different kinds.  From $G$ we compute the $\rng (G)$, the
mountain graph $\mg(G)$ and the line parent hierarchy $\lp(G))$ of the
essential landscape. For each $d$ and $|M_2|$, we repeat this procedure
20 times and display means.  The results can be found
in~\cref{fig:random}.  The following facts stand out.
\begin{itemize}
\item The density of both the $\rng$ and the mountain graph of the
  $\rng$ are growing in a significant smaller pace than the density of
  $G$. Considering the case $|M_2|=100,000$ it is remarkable, that, when
  $G$ has a density of nearly $1$, the density of both other graphs
  are still under $0.3$.
\item The characteristic points for describing the resulting mountain
  landscape build indeed a subset that is remarkably smaller than the
  vertex size of the biggest component of $G$.
\item Considering the second row, it stands out that for very high
  densities of nearly $1$ of the original bipartite network, the
  density and thus the amount of edges of the RNG start to rise
  rapidly. We assume that this is driven by the case, that, if the
  bipartite graph is nearly complete, there will be a large amount of
  vertex pairs with the same neighbor set in the bipartite
  graph. Thus, nearly all shortest path distances are equal and just
  very few edges will be discarded. In consequence, the mountain graph
  is built from a nearly complete graph where nearly all edges have
  similar weights. Thus, there will be only a small amount of peaks
  and the mountain graph is nearly vanishing.
\end{itemize}

Note, that we use a height function that is directly derived from the graph.
Such height functions are indeed reasonable.  For example, the amount of
followers of Twitter users in co-follower graphs is indeed a useful indicator of
their importance.

\section{Conclusion and Future Work}
\label{sec:conclusion}

In this work, we showed how the notions of peaks, cols and line parents, which
are originally designed to characterize connections and hierarchies between
mountains, can be adapted to networks. We discussed how these notions can be
used to identify important vertices and meaningful connections and hierarchies
between them. Our method further benefits from a novel preprocessing procedure
that removes unimportant edges without hurting the connectivity of the network.
This preprocessing step is based on relative neighborhood graphs which were
originally invented to connect data points in two-dimensional euclidean spaces.

Our experiments indicate  that our method finds dependencies and
hierarchies from the original network that are remarkably smaller than the
original graph and therefore can enhance the comprehension of real-world social
networks.

Future work will investigate the application to further kinds of
networks such as friendship networks on Facebook. As some of these
networks may be unweighted, the question arises on how to use our RNG
procedure in this case. On the other hand, it would be interesting to involve temporal
aspects in networks. How does the line parent hierarchy of 
social networks change over time?

\bibliographystyle{splncs04}
\bibliography{literature}

\end{document}